\documentclass
[nofootinbib,superscriptaddress,notitlepage,a4paper,twocolumn]{revtex4-1}%
\usepackage{amssymb}
\usepackage{amsfonts}
\usepackage{amsmath}
\usepackage{graphicx}
\usepackage{color}
\usepackage{version}
\usepackage{numinsec}%
\providecommand{\U}[1]{\protect \rule{.1in}{.1in}}

\includeversion{Commentskip}
\includeversion{Commentdel}
\includeversion{Commentnew}

\begin{document}
\title{Self-energy correction to dynamic polaron response}
\author{Dries Sels}
\email{Corresponding author: dries.sels@uantwerpen.be}
\author{Fons Brosens}
\email{fons.brosens@uantwerpen.be}
\affiliation{Physics Department, University of Antwerp, Universiteitsplein 1, 2060
Antwerpen, Belgium}

\begin{abstract}
We present the first order self-energy correction to the linear response
coefficients of polaronic systems within the truncated phase space approach
developed by the present authors. Due to the system-bath coupling, the
external pertubation induces a retarded internal field which dynamically
screens the external force. Whereas the effect on the mobility is of second
order, dynamical properties such as the effective mass and the optical
absorption are modified in first order. The Fr\"{o}hlich polaron is used to
illustrate the results.

\end{abstract}
\maketitle

\section{Introduction}

In a previous paper~\cite{cit:SelsBrosensTruncate} we presented an
approximate, however systematically improvable, truncation method to derive
the linear response coefficients from the quantum Liouville equation for the
reduced Wigner function~\cite{cit:SBreducedLiouville2013} of polaronic
systems. The paper mainly addressed the discrepancy between the mobility of
the Fr\"{o}hlich
polaron~\cite{cit:Frohlich1937,cit:Frohlich1950,cit:Frohlich1954} proposed by
Feynman \textit{et al.}~\cite{cit:FHIP1962}~ (hereafter referred to as FHIP)
and Kadanoff \cite{cit:FHIP1962}. It was shown how a slight modification to
each of the two methods, which accounts for their discrepancy and amends their
problems, makes them compatible with the presented truncation method.
Moreover, the new result turned out to be in agreement with a prediction made
by Los'~\cite{cit:LOS}.

In the present paper we concentrate on the dynamic response properties of the
polaron system
\begin{multline}
H=\frac{\mathbf{p}^{2}}{2m}-e\mathbf{E(}t\mathbf{)}\cdot \mathbf{x}%
+\sum_{\mathbf{k}}\hbar \omega_{\mathbf{k}}\left(  b_{\mathbf{k}}^{\dagger
}b_{\mathbf{k}}+\frac{1}{2}\right) \label{eq:Hamiltonian}\\
+\sum_{\mathbf{k}}\left[  \gamma(\mathbf{k})\exp \left(  i\mathbf{k\cdot
x}\right)  b_{\mathbf{k}}^{\dagger}+\gamma^{\ast}(\mathbf{k})\exp \left(
-i\mathbf{k\cdot x}\right)  b_{\mathbf{k}}\right]  ,
\end{multline}
where $\left(  \mathbf{x,p}\right)  $ represents the particle which is coupled
to some bosonic field $b_{\mathbf{k}}$ in a isotropic translational invariant
way, i.e. $\gamma(\mathbf{k})=\gamma(\left \vert \mathbf{k}\right \vert )$ and
$\omega_{\mathbf{k}}=\omega_{\left \vert \mathbf{k}\right \vert }$. The
particle, which we consider to be charged, is subject to a small time
dependent electric field $\mathbf{E(}t\mathbf{)}$ and we are concerned with
finding the time dependent response of the system in the form of the
conductivity
\begin{equation}
\mathbf{J}(t)=\int_{-\infty}^{t}\sigma(t-s)\mathbf{E(}s\mathbf{)}%
\mathrm{d}s=\frac{e}{m}\int \mathbf{p}f\left(  \mathbf{p},t\right)
\mathrm{d}\mathbf{p,} \label{eq:Generic:definesigma}%
\end{equation}
where $f\left(  \mathbf{p},t\right)  $ is the reduced momentum distribution of
the system.

\section{Self-energy correction and dynamical screening}

Following Ref.~\cite{cit:SelsBrosensTruncate} we wish to derive an equation of
motion for the current density under the assumption that the bosonic field
$b_{\mathbf{k}}$ was initially in thermal equilibrium. It follows from
definition (\ref{eq:Generic:definesigma}) and from the relevant equations
(2.5-2.7) of Ref.~\cite{cit:SelsBrosensTruncate} that the current density
satisfies%
\begin{multline}
\frac{d\mathbf{J}(t)}{dt}-\frac{e^{2}}{m}\mathbf{E(}t\mathbf{)}=-\sum_{k}%
\frac{2\left \vert \gamma(\mathbf{k})\right \vert ^{2}}{\hbar}\mathbf{\mathbf{k}%
}\frac{e}{m}\int \mathrm{d}\mathbf{p}\int_{-\infty}^{t}dt^{\prime}\\
\times \left \{
\begin{array}
[c]{c}%
n_{B}(\omega_{k})\sin \left[  \left(  \frac{\mathbf{\hbar \mathbf{k}}^{2}}%
{2m}-\omega_{k}\right)  (t-t^{\prime})\right] \\
+\left(  n_{B}(\omega_{k})+1\right)  \sin \left[  \left(  \frac{\mathbf{\hbar
\mathbf{k}}^{2}}{2m}+\omega_{k}\right)  (t-t^{\prime})\right]
\end{array}
\right \} \\
\times \sin \left(  \mathbf{k\cdot}\left[  \frac{\mathbf{p}}{m}(t-t^{\prime
})+\int_{t^{\prime}}^{t}\int_{t^{\prime}}^{\tau}\frac{e\mathbf{E}(s)}%
{m}\mathrm{d}s\mathrm{d}\tau \right]  \right)  f\left(  \mathbf{p},t^{\prime
}\right)  .
\end{multline}
At low temperature and for weak coupling, the momentum distribution function
is assumed to be peaked around a small average value of $\mathbf{p}$, because
the perturbation $\mathbf{E}$ is assumed to be weak. It thus seems reasonable
to expand the sine function. Truncating the expansion up to first order
results in%
\begin{multline}
\frac{\partial}{\partial t}\mathbf{J}(t)+\int_{-\infty}^{t}\chi(t-s)\mathbf{J}%
(s)\mathrm{d}s=\frac{e^{2}}{m}\mathbf{E(}t\mathbf{)}\\
\mathbf{-}\frac{e^{2}}{m}\int_{-\infty}^{t}\kappa(t-s)\mathbf{E}%
(s)\mathrm{d}s,
\end{multline}
where we have adopted the notation of \cite{cit:SelsBrosensTruncate}, such
that memory function $\chi$ of the system is given by
\begin{multline*}
\chi(t)=t\sum_{k}\frac{2\left \vert \gamma(\mathbf{k})\right \vert ^{2}}{3\hbar
}\frac{\mathbf{\mathbf{k}}^{2}}{m}\\
\times \left \{
\begin{array}
[c]{c}%
n_{B}(\omega_{k})\sin \left(  \left[  \frac{\hbar \mathbf{k}^{2}}{2m}-\omega
_{k}\right]  t\right) \\
+\left(  n_{B}(\omega_{k})+1\right)  \sin \left(  \left[  \frac{\hbar
\mathbf{k}^{2}}{2m}+\omega_{k}\right]  t\right)
\end{array}
\right \}  ,
\end{multline*}
and the polarizability $\kappa$ becomes%
\begin{equation}
\kappa(t)=t\int_{t}^{\infty}\mathrm{d}\tau \frac{\chi(\tau)}{\tau}.
\label{eq:Polarizability_eta}%
\end{equation}
Consequently, according to Eq.~(\ref{eq:Generic:definesigma}), the Laplace
transform $%
\mathcal{L}%
\left(  \sigma,\Omega \right)  $ of the conductivity satisfies%
\begin{equation}%
\mathcal{L}%
\left(  \sigma,\Omega \right)  =\frac{e^{2}}{m}\frac{1-%
\mathcal{L}%
\left(  \kappa,\Omega \right)  }{\Omega+%
\mathcal{L}%
\left(  \chi,\Omega \right)  }. \label{eq:ConductivityLaplace_notresummed}%
\end{equation}
A more accurate conductivity can be found using a resummation argument similar
to that in~\cite{cit:SBreducedLiouville2013}, which yields%
\begin{equation}%
\mathcal{L}%
\left(  \sigma,\Omega \right)  \approx \frac{e^{2}}{m}\frac{1}{\Omega+\left(
\Omega%
\mathcal{L}%
\left(  \kappa,\Omega \right)  +%
\mathcal{L}%
\left(  \chi,\Omega \right)  \right)  }. \label{eq:ConductivityLaplace}%
\end{equation}
The resummation approximately takes into account that the proper
polarizability $\kappa$ depends on the response $\sigma$ of the system itself.
Of course at very small coupling it would not matter. It ought to be clear
that expression (\ref{eq:ConductivityLaplace}) for the conductivity reduces to
expression (3.4) in Ref.~\cite{cit:SelsBrosensTruncate} under the condition
that $\Omega%
\mathcal{L}%
\left(  \kappa,\Omega \right)  =0.$ Consequently, for every finite $\kappa_{0}=%
\mathcal{L}%
\left(  \kappa,0\right)  $ the dc-conductivity is equal to the dc-conductivity
discussed in \cite{cit:SelsBrosensTruncate}. Corrections to the mobility due
to dynamical screening are thus of second order. However, consider a Taylor
expansion around $\Omega=0$ of the memory function $\chi$ and the
polarizability $\kappa$
\begin{align*}%
\mathcal{L}%
\left(  \chi,\Omega \right)   &  =\chi_{0}+\chi_{1}\Omega+O(\Omega^{2}),\\%
\mathcal{L}%
\left(  \kappa,\Omega \right)   &  =\kappa_{0}+\kappa_{1}\Omega+O(\Omega^{2}).
\end{align*}
Then we find the low energy optical absorption%
\[
\operatorname{Re}\left[
\mathcal{L}%
\left(  \sigma,i\omega \right)  \right]  \approx \frac{e^{2}\pi}{m\left(
1+\chi_{1}+\kappa_{0}\right)  }\left[  \frac{1}{\pi}\frac{\gamma}{\omega
^{2}+\gamma^{2}}\right]  ,
\]
where $\gamma=\chi_{0}\left(  1+\chi_{1}+\kappa_{0}\right)  ^{-1}.$ This
implies that the effective mass is given by%
\begin{equation}
\frac{m^{\ast}}{m}=\left(  1+\chi_{1}+\kappa_{0}\right)  =1+\frac{\chi_{1}}%
{2}. \label{eq:effectiveMass}%
\end{equation}
The latter equality immediately follows from the definition
(\ref{eq:Polarizability_eta}) of $\kappa(t)$ in terms of $\chi(t).$ In
contrast to the mobility, which remains unchanged to first order, the
effective mass of the system is significantly altered by dynamical screening.
In fact the relative change in the mass is only half of the change without
dynamical screening. It should be noted that effective mass is a dynamical
quantity and it depends on the entire spectral function through the
polaron-f-sum rule~\cite{cit:DLVRsumrule} by Devreese \textit{et al..
}Consequently, a redistribution of spectral weight must accompany the change
in effective mass. Let us illustrate this with the Fr\"{o}hlich
polaron~\cite{cit:Frohlich1937,cit:Frohlich1950,cit:Frohlich1954}.

\section{Fr\"{o}hlich polaron\label{sec:Frohlich}}

We take $\hbar=m=1.$ For the Fr\"{o}hlich polaron we moreover consider
$\omega_{k}=\omega_{LO}=1$ and $\left \vert \gamma \left(  k\right)  \right \vert
^{2}=2\sqrt{2}\pi \alpha V^{-1}k^{-2}.$ According to
Ref.~\cite{cit:SelsBrosensTruncate}, the Laplace transform of $\chi$ is%
\begin{equation}%
\mathcal{L}%
\left(  \chi,\Omega \right)  =\frac{\alpha}{3\sqrt{\Omega^{2}+1}}\left(
\begin{array}
[c]{c}%
\left(  2n_{B}+1\right)  \sqrt{\sqrt{\Omega^{2}+1}+\Omega}\\
-\sqrt{\sqrt{\Omega^{2}+1}-\Omega}%
\end{array}
\right)  . \label{eq:Frohlich:chi_Laplace}%
\end{equation}
By expanding around $\Omega=0$ one readily finds $\chi_{1}=\alpha \left(
n_{B}+1\right)  /3,$ such that the zero temperature effective mass is
\[
\frac{m^{\ast}}{m}=1+\frac{\alpha}{6},
\]
in agreement with standard weak coupling theories,\ for which we refer
to~\cite{cit:AlexandrovDevreese2010,cit:DevreeseVarenna}. The $T=0$ Laplace
transform $%
\mathcal{L}%
\left(  \kappa,\Omega \right)  $ of the polarizability is given by the
following integral%
\[%
\mathcal{L}%
\left(  \kappa,\Omega \right)  =\frac{4\alpha}{3\pi}\int_{-\infty}^{\infty
}\mathrm{d}u~u^{2}\frac{\Omega^{2}-\left(  u^{2}+1\right)  ^{2}}{\left(
u^{2}+1\right)  \left(  \left(  u^{2}+1\right)  ^{2}+\Omega^{2}\right)  ^{2}%
},
\]
which can readily be done by using Cauchy's residue theorem, which yields%
\[%
\mathcal{L}%
\left(  \kappa,\Omega \right)  =\frac{\sqrt{2}\alpha}{3\Omega^{2}}\frac
{\Omega^{2}+2+2\sqrt{\Omega^{2}+1}}{\sqrt{\Omega^{2}+1}\sqrt{\sqrt{\Omega
^{2}+1}+1}}-\frac{4\alpha}{3\Omega^{2}}.
\]

\begin{figure}
[ptb]
\begin{center}
\includegraphics[
height=6.4581cm,
width=8.6042cm
]%
{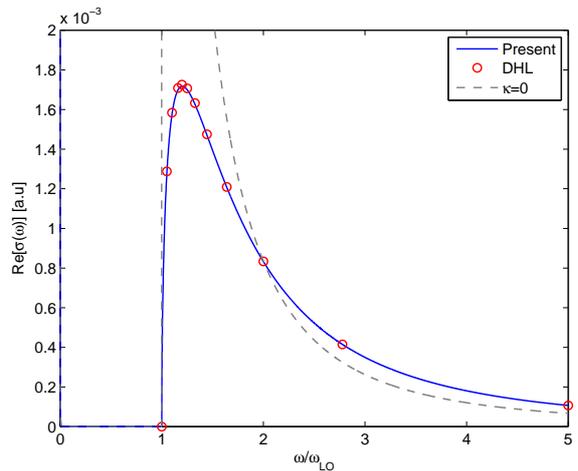}%
\caption{Optical absorption coefficient for Fr\"{o}hlich polaron at $T=0$ for
$\alpha=0.01.$ The full blue line represent the present result, the dashed
gray line ($\kappa=0$) would be the result obtained without
screening~\cite{cit:SelsBrosensTruncate} and the red circles (DHL) is a
perturbative result by Devreese \textit{et al}. \cite{cit:DHL}.}%
\label{fig:Compareweakalpha}%
\end{center}
\end{figure}

\begin{figure}
[ptb]
\begin{center}
\includegraphics[
height=7.3016cm,
width=8.6042cm
]%
{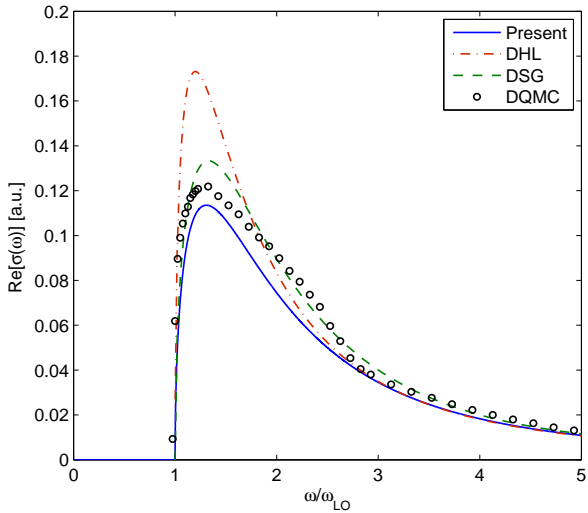}%
\caption{Optical absorption coefficient for Fr\"{o}hlich polaron at $T=0$ for
$\alpha=1.$ The full blue line represent the present result, the dashed red
line (DHL) is a perturbative result by Devreese \textit{et al}. \cite{cit:DHL}%
. Furthermore, the dashed green line (DSG) is a variational result due to
Devreese \textit{et al.}~\cite{cit:DSG}\textit{ }and the circles (DQMC) show a
numerical result due to Mishchenko \textit{et al}.~\cite{cit:DQMC}. DSG and
DQMC data copied with permission of the authors from~\cite{cit:KliminDevreese}%
.}%
\label{fig:comparealpha1}%
\end{center}
\end{figure}

For $\Omega=0$ this indeed results in $\kappa_{0}=$ $%
\mathcal{L}%
\left(  \kappa,0\right)  =-\alpha/6.$ The $T=0$ optical absorption is depicted
in Fig. (\ref{fig:Compareweakalpha}) and Fig. (\ref{fig:comparealpha1}) for
$\alpha=0.01$ and $\alpha=1$ respectively.
Fig.~(\ref{fig:Compareweakalpha}) clearly shows the effect of dynamical
screening on the optical absorption. When the effect of the induced electric
field is ignored, i.e. $\kappa=0,$ the absorption becomes more singular near
the absorption threshold and the high frequency absorption is slightly
reduced. As implied by the polaron-f-sum rule~\cite{cit:DLVRsumrule}, the
total absorption for $\omega>\omega_{LO}$ is smaller when dynamical screening
is taken into account. In agreement with the polaron-f-sum rule, the reduction
of the total absorption beyond threshold reduces the relative change in the
mass by a factor 2. For comparison Fig. (\ref{fig:Compareweakalpha}) also
shows a weak coupling result due to Devreese \textit{et al}.~\cite{cit:DHL}
(hereafter referred to as DHL). Their result is perturbative in $\alpha$ and
thus becomes exact for $\alpha \rightarrow0.$ For $\alpha=0.01$ their result is
indistinguishable from the present result, which implies the present
truncation scheme correctly predicts the weak coupling optical absorption. For
$\alpha=1$ we show the absorption spectrum in Fig.~(\ref{fig:comparealpha1}).%
At this point there is a clear distinction between the present approach and
the DHL result. We therefore compare the result with the absorption obtained
from a diagrammatic quantum Monte Carlo calculation~\cite{cit:DQMC} which
should give numerically exact answers for all $\alpha.$ Although the present
result is distinguishable from the Monte Carlo calculation, it is clearly more
accurate than the perturbative result of DHL. Moreover, the present result is
remarkably close to the nonperturbative method presented in Ref.
\cite{cit:DSG}. The method, due to Devreese \textit{et al}.~\cite{cit:DSG},
employs the impedance function approximation of FHIP~\cite{cit:FHIP1962}. The
method is thus nonperturbative in the sense that no expansion in the coupling
constant is assumed.

\bigskip

\section{Conclusion}

In conclusion we have presented the first order self-energy correction to the
linear response coefficients of polaronic systems within the truncated phase
space approach developed by the present authors in
\cite{cit:SelsBrosensTruncate}. It is shown how the change of the self-energy
due to the external perturbation induces an internal field. The first order
correction thus comes in terms of a dynamic polarizability $\kappa.$ It is
shown that the relative change in the effective mass is only half of the
change without dynamical screening. Consequently, a significant amount of
spectral weight must be moved to the central peak. Explicit expressions for
the conductivity of the Fr\"{o}hlich polaron are obtained. The results are
shown to be in agreement with standard weak coupling theories for $\alpha
\ll1.$ Comparing with numerically exact data, we found that the present
approach significantly improves on the standard weak coupling perturbation
theory and extends the validity up to $\alpha \approx1.$

\begin{acknowledgments}
The authors thank J.T. Devreese for many stimulating discussions, in
particular on the polaron-f-sum rule and for providing numerical data on the
$\alpha=1$ absorption.
\end{acknowledgments}

\end{document}